\documentclass[12pt,a4paper]{article}
\usepackage[utf8]{inputenc}
\usepackage{amsmath}
\usepackage[table]{xcolor}
\usepackage{xcolor}
\usepackage{multirow}
\definecolor{myOrange}{RGB}{44,14,14}
\usepackage{amsfonts}
\usepackage{afterpage}
\usepackage{amssymb}
\usepackage{array}
\usepackage{tabularx}
\newcolumntype{P}[1]{>{\centering\arraybackslash}p{#1}}%for vertical aligment in table
\newcolumntype{M}[1]{>{\centering\arraybackslash}m{#1}}%for vertical aligment in table

\usepackage{dsfont}
\usepackage{hyperref}
\usepackage{relsize}
\usepackage{graphicx}
\usepackage{graphics}
\usepackage{grffile}
\usepackage[section]{placeins}
\usepackage{sectsty}
\usepackage{cancel}
\sectionfont{\fontsize{15}{15}\selectfont}
\subsectionfont{\fontsize{13}{15}\selectfont}
\usepackage{marginnote}
\usepackage[obeyFinal,textsize=tiny,color=yellow]{todonotes}
\usepackage{authblk} %including authors afiliations in main page
\usepackage{lmodern,textcomp}
\usepackage[official]{eurosym}
\usepackage{float}
\restylefloat{table}
\usepackage{siunitx}
\usepackage{lscape}
\usepackage{caption}
\newcolumntype{P}[1]{>{\centering\arraybackslash}p{#1}}
\usepackage{enumerate}
\usepackage[style=authoryear-icomp,maxbibnames=9,maxcitenames=2,uniquelist=false, backend=biber, sortcites, natbib]{biblatex}

\defbibenvironment{bibliography}
  {\begin{enumerate}}
  {\end{enumerate}}
  {\item}

\bibliography{library, library}
\usepackage[english]{babel}
\bibdata{library}
\newcommand{\note}[1]{\marginpar{\begin{flushleft}
\end{flushleft}}}

\title{BEYOND FISHING: THE VALUE OF MARITIME CULTURAL HERITAGE IN GERMANY}
\author[]{Emily Quiroga\thanks{Corresponding author: emily.quiroga-gomez@uni-hamburg.de}}
\affil[]{Research Unit Sustainability and Climate Risk, University of Hamburg, Grindelberg 5, 20144 Hamburg, Germany}
\date{}

\allowdisplaybreaks

\begin{document}

\maketitle

\begin{abstract}
The importance of maritime heritage in providing benefits such as a sense of place and identity has been widely discussed. However, there remains a lack of comprehensive quantitative analysis, particularly regarding monetary valuation and its impact on people's preferences. In this study, I present the results of a choice experiment that assesses the value of the maritime cultural heritage associated with shrimp fishing through seafood consumption preferences in Germany. Additionally, I investigate people's attitudes toward cultural heritage and examine how these attitudes affect their stated preferences. I find that these attitudes are significantly stronger in towns where local fishermen led a prominent awareness campaign on fishing culture during the study period. Moreover, I observe a positive willingness to pay for a cultural heritage attribute in shrimp dishes, which varies depending on individuals' attitudes toward cultural heritage.

\end{abstract}

\newpage
\section{Introduction}
Fishing is a practice dating back to ancient times. It contributes as a source of food, income and pride. In the European Union (EU), the fishing activity provides 124.000 direct jobs, with nearly the half belonging to the small-scale coastal fleet. Fishing activity also contributes to the personal protein intake, where an average person consumes 3.3Kg more sea food than the world average \citep{STECF2022}. Besides an economic benefit, fishing, particularly in coastal areas, has an inherit cultural heritage characterized by the creation of a sense of place in terms of place attachment and cultural-social memory for residents \citep{Khakzad2016}. The fishing cultural heritage is a public good characterized by fishing traditions, maritime cultural landscape and traditional waterfronts. These elements can create place attachment for both residents and visitors in these areas \citep{Khakzad2016}\footnote{Place attachment refers to connections to physical and social settings that provide social and psychological benefits to residents and visitors in these fishing areas \citep{Brown2003}. Cultural heritage, in particular, is a form of asset belonging to an individual, community, a region, a nation or to humankind as a whole \citep{Throsby2007}.}.
Maritime cultural heritage maintains a sense of place in fishing communities and contribute to preserve the socio-cultural memory, as well as social and psychological benefits  \citep{Khakzad2016,Duran2015,Urquhart2013, Martino2023, Xiao2023}.  

In Germany, this maritime cultural heritage is at risk. The brown shrimp fishery, currently the most important coastal fishery, experience a constant decrease in the number of vessels over the last two decades from 250 to 145 in 2024\footnote{Data provided by the Th\"{u}nen Institue}. Despite not being constrained by quotas this fishery is subject to the European Common Fishery Policy (CFP) regulations, and  national restrictions such as mesh size and the number of licenses issued \citep{Doring2020}. The adaptation of this fishery over the decades is remarkable given historic and economic facts such as the World War II, the reunification of Germany, the declaration of Exclusive Economic Zones (EEZ) after 1977 \citep{Schacht2023}, the increase in fuel prices, increasing imports of products at a lower price, and abrupt changes in demand among others \citep{STECF2022}. This fishery embeds a set of cultural traditions and knowledge that has remain despite these multiple challenges.

The public good nature of maritime cultural heritage implies that some elements would not survive without some form of collective action, as the market does allocate resources properly, with a provision lower than socially optimal \citep{Duran2015}. Yet, the challenge of determining the optimal provision together with insufficient data, particularly on the demand side, may result in the under-provision of this public good \citep{Throsby2007}. Given the importance of maritime cultural heritage and the risk of its disappearance, this study aims to assess the Willingness To Pay (WTP) of residents and non-residents for shrimp cultural heritage in Germany. I analyse the maritime heritage aspect of sea food choices encompassing vibrant traditional working waterfronts with operational boats and bustling fish markets. These elements significantly impact people's decisions to consume seafood \citep{SYMES2009, Khakzad2016}.

This paper adds to the literature by exploring the link between food consumption and maritime cultural heritage. To my knowledge, only \citet{Martino2023} explored the link between food consumption and cultural heritage in Scotland. This paper is the first study in Germany assessing consumers preferences for maritime cultural heritage. I also contribute by examining people's attitudes that influence consumers preferences towards maritime cultural heritage. I implement a modified survey based on the attitudinal scale developed by \citet{Choi2007}. This scale measures cultural attitudes of people using factor and hierarchical cluster analysis. I identify aspects that influence pro-cultural attitudes towards maritime heritage. 

I also exploit a campaign that unfolded on-site during the study period. The campaign promoted by local shrimp fishers, aimed  to raise awareness about the shrimp cultural heritage.  The campaign was more visible in two of the four towns where this study was conducted. I examine the correlation of the awareness-raising campaign towards shrimp cultural heritage on peoples attitudes towards cultural heritage and WTP. I find a  significant correlation among the presence of the campaign with  higher pro-heritage attitudes, and an increased WTP for maritime cultural heritage. The results show that people have a positive WTP for shrimp cultural heritage but also for a local, fresh and sustainable shrimp dish.  In the next section I describe the German North Sea shrimp fishery, in section three I describe the methodology, in section fourth I present the results, in the fifth section I show  a discussion, and the final section offers the conclusion.

\section{North Sea shrimp maritime cultural heritage}

The Brown shrimp is mostly harvested by Germany and the Netherlands, which together account for 79\% of the EU brown shrimp production \citep{Goti-Aralucea2021}. These fisheries are organized as producer organizations grouped either by size of the vessel or the geographical location \citep{Goti-Aralucea2021}. The majority of these fisheries are certified by the Marine Stewardship Council (MSC), which now serves as a management system for the shrimp fishery. This certification was initially developed for a consortium of German fisheries and now the plan encompasses 421 vessels from the Netherlands, Germany, Denmark, and Belgium, detailing specifications for gear, mesh size, and beam length, as well as guidelines for reducing catch per unit of effort in accordance with the criteria of the International Council for the Exploration of the Sea (ICES) \citep{Addison2023, Goti-Aralucea2021}.

In the last years brown shrimp fisheries face increasing regulations and socio-economic pressures that  endanger their existence. The increasing area for Offshore Wind Farms (OWF) leave fisheries with less available space to fish. In the Wadden Sea 40\% of the coastal area is free from fishing \citep{Stelzenmuller2021}. Moreover, higher temperatures could cause migration of the species to other areas \citep{Schulte2020}, and the recently COVID pandemic drop drastically the prices and landings decreased \citep{STECF2022}. It is challenging to maintain profitability despite some efficiency factors such as, externalization of shelling and marketing tasks, together with capital cost reduction \citep{Goti-Aralucea2021}. Various resilience strategies are contemplated to improve the business, including mechanical shelling, instead of the outsourcing shelling in Morocco, and internalization of the marketing and sale of the shrimps. Further strategies aim to integrate greater diversity within both the fleet and administrative teams to foster innovation within the shrimp business \citep{Goti-Aralucea2021}.

\begin{figure}[!h]
\centering
% \begin{minipage}{\textwidth}
  \includegraphics[width=0.8\textwidth]{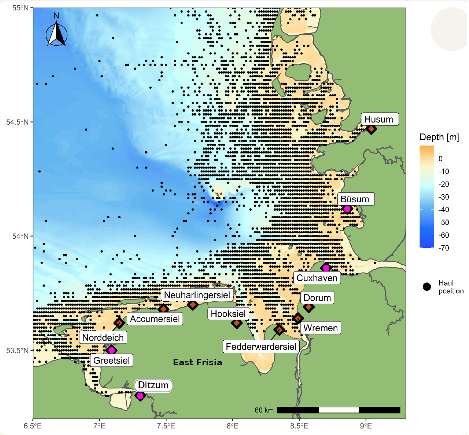}
  \caption[]{Brown shrimp ports in Germany. The pink dots represent the ports (towns) where this study was carried out. Figure based on the study by \citet{Goti-Aralucea2021}.} 
  \label{fig:mapDCE}
 % \end{minipage}
\end{figure}

Besides the mentioned pressures on fisheries, during March of 2023 in B\"{u}sum, the agriculture ministry announced an action plan proposed by the European Commission which includes the ban of the fishing method of bottom trawls, arguing that it damages the seabed. Although the proposed plan still requires a negotiation which each member state, the mayor of Greetsiel mentioned that this measure could lead to the disappearance of cultural heritage, tourism would be negatively affected, and business could lose their existence. Fishers associations claim that this regulation could mean an end to the traditional fishing method \citep{NDR2023, NDR20232}. 
After the announcement, fishers started a campaign to support the existence of the brown shrimp trawlers. The campaign consists of symbols such as crosses symbolizing the death of the vessels, including a letter explaining the causes they support (See annex figures \ref{fig:photoCampaign} and \ref{fig:photoLetter}). Fishers mention that the prohibition of the fishing method and the decreasing space available to fish would mean the end of brown shrimp fishing, fishing companies, and the tourism industry. This letter was displayed in most restaurants and touristic key points such that it was highly visible to locals and tourists. The campaign was more visible in the towns (ports) of Greetsiel and Ditzum, due to the small touristic area in comparison to Cuxhaven and B\"{u}sum (See figure \ref{fig:mapDCE}).

%MAP DCE

The consequences of the possible disappearance of the Brown Shrimp fishery could have significant economic implications. It is the the most important coastal fishery for Germany in economic terms, given by its large volume and high prices that yield the highest revenues of all coastal fleets \citep{Goti-Aralucea2021}. However, the consequences of loosing this fishery go beyond the mere act fishing, encompassing potential impacts on cultural heritage.  Fisheries in Germany started before the Middle Ages \citep{Doring2020}, and the tradition of fishing Shrimp is one of the oldest cultural fishing techniques in the North Sea \citep{OSTFRIESKRIMI2023}. However, the current form of fishing with Beam Trawls is not older than approximately 120 years old. The maritime cultural heritage includes the historical ports, the active vessels, and the traditions and architecture surrounding the fishing activity. The figure \ref{fig:mapDCE} shows important ports where the German brown shrimp is landed and the fishing area. In Greetsiel, for instance, the port is older than 600 years, making the town still preserving the “magic of old times"\citep{Greetsiel2023}.  The North Sea shrimp is also one of the symbols of the East Frisia region (See figure \ref{fig:mapDCE}), as it contributes to build the identity of many people from the coastal region of the North Sea.

The goods and services provided by the brown shrimp fishery go beyond providing a source of protein, the maritime cultural heritage involved in the fishing activity contributes to the individual and collective well-being. Hence, the estimation of the economic value of the cultural heritage as a non-market good becomes a relevant part of cultural policies \citep{Mourato2002,Ready2002}. However, for the services provided by the cultural heritage there is not an associated price due to the non-market nature. This study contributes to obtain the economic value of the brown shrimp cultural heritage using a stated preferences methodology where individuals determine the utility of a good based on its attributes.

\section{Methodology}

I conducted a survey and a Discrete Choice Experiment (DCE) to identify the value of the cultural heritage of the shrimp fishery. DCE is an economic valuation method that  through a simulated market scenario enables the analysis of stated preferences for non-markets goods.   This methodology is extensively used in the context of cultural heritage, for instance, museums or places that are unique by definition \citep{Choi2010,Bedate2004,Bertacchini2011}. Recently, DCE has been applied in maritime cultural heritage studies \citep{Xiao2023, Martino2023, Tanner2021}, with \citet{Duran2015} being the lead to use this method in the field.  

I conducted a survey in four towns with the highest number of shrimp landings in Germany (Figure \ref{fig:mapDCE}). The ports with the highest average proportion of landings in the last seven years, in descending order, are: B\"{u}sum (26\%), Cuxhaven (12\%), and Greetsiel (10\%). Following Greetsiel, there are six towns with similar shares of shrimp landings (ranging between 3.5\% and 4.5\%). Based on expert guidance, Ditzum was selected as the fourth town due to its maritime cultural significance, with a landings share of 3.9\% \footnote{The guidance was based on experts from the research institute of the German Ministry of Food, Agriculture, and Consumer Protection (Th\"{u}nen Institute)}. The selected towns have few locals, and they live mainly from tourism. The local population in these places is 570 in Ditzum, 1.410 in Greetsiel, 4.876 in B\"{u}sum and 48.562 in Cuxhaven \footnote{Data from \citet{CityPopulation2023}}. The quantity of visitors exceeds by a high magnitude the quantity of locals. In 2021, Emden, region where Ditzum is located, had 101.167 visitors. Krummhörn, the region to which Greetsiel belongs, had 76.039 tourists. B\"{u}sum received 228.036 and Cuxhaven 358.728 tourists \footnote{Statistics for 2021 data from \citet{Stadtistik2021}}.

This study was conducted in August of 2023 for twenty consecutive days. In that period, a fishing campaign was underway in the harbour and historic centres of the towns. The visibility of the campaign for tourists varied between towns and  depended on the spatial connection of the historic center to the harbour and the availability of an open sea area (beach). In Ditzum and Greetsiel, the location of the historical center, harbour, and  tourists routes are interconnected, attracting visitors primarily interested in nature, fishing and local gastronomy. Due to the compact size of their historic centres and the absence of an open sea area, tourist routes are concentrated around key locations where campaign advertisements were prominent (See figure \ref{fig:photoCampaign}). In contrast,  B\"{u}sum and Cuxhaven have an open sea area (beach) that often serve as the main attraction for visitors. In Cuxhaven, for instance, the historic center is 5km away of the beach, and the harbour is disconnected from both the city center and the beach.  These factors contributed to the campaign being more visible in Ditzum and Greetsiel than in B\"{u}sum and Cuxhaven.

\subsection{Survey design}

I designed and implemented a face-to-face survey involving 409 individuals to assess preferences towards the shrimp cultural heritage. The recruitment process was random to tourist and residents in these areas. The survey comprises five sections (1) Background information of the respondent. (2) Attitudinal scale towards shrimp cultural heritage. (3) Connection with fishery (4) The choice experiment and (5) socio-demographic information.

The first section corresponds to the background of the interviewer, reasons to travel, the federal state where they come from, and the duration of their stay. The second section includes an attitudinal sale to asses shrimp cultural heritage values. I adapted \citet{Choi2007}'s attitudinal scale to the shrimp cultural heritage context, and added important aspects mentioned in \citet{Martino2023} regarding socio-economic dimensions. The attitudinal scale comprises: (a)  Intercommunity and intergenerational linkages (b) Recognition of diverse cultural values (c) Awareness of cultural loss (d) Preservation of traditions and customs (e) Economic, environmental and social dimensions. All statements are assessed in a scale from 1 (strongly disagree) to 5 (strongly agree). A the beginning of the survey participants received an explanation about the meaning of shrimp marine cultural heritage \footnote{The cultural heritage definition used is derived from \citet{Khakzad2016}. It refers to tangible and intangible values associated with shrimp fishing.}. The third section addresses the reasons for visiting or living in the town, the connections with fishing and importance of the harbour. The four section corresponds to the DCE and the fifth addresses socio-demographic information.  The survey is available in the \autoref{sec:appendix1}.

\subsection{Discrete choice experiment}

The DCE estimates the price premiums consumers are willing to pay for characteristics (attributes) of a shrimp meal with a side. It models consumer preferences by presenting respondents with various combinations of these attributes, prompting them to make trade-offs among alternatives. Consumers then select the alternative that maximizes their utility, allocating their income to attain the optimal combination of attributes. This process enables the inference of Willingness To Pay (WTP) for changes in those attributes, providing a monetary measure of the change in utility levels as consequence of a change in the characteristics of this good \citep{Mariel2021}. This methodology enables a relevant comparison between various goods and state an unbiased metric for gauging preferences over attributes \citep{Mariel2021}.

In this study a shrimp meal with a side is the vehicle through which the value of cultural heritage is assessed.  The shrimp meal attributes that participants evaluate in the CE are derived from previous studies \citep{Duran2015,Martino2023,Verbeke2016} and are described as follows: 1) \textit{Origin of the shrimp}: whether it is locally produced or imported; 2) \textit{Processing of the fish}: whether it is fresh or frozen; 3) \textit{Harvesting process}: distinguishing between small-scale inshore fishing conducted by local traditional vessels or foreign vessels. 4) \textit{Certification}: referring to sustainable fishing labels as an environmental protection measure, comparing certified and non-certified fisheries \citep{Cerjak2014}; 5) \textit{The heritage: } encompassing the visual aspect of inshore shrimp active fishing, including the cultural experience of a visible active fishing boats versus limited access due to waterfront development for residential and non-fishing commercial purposes; and 6) \textit{Payment vehicle: }used to calculate monetary trade-offs concerning a shrimp meal with a side. The payment attribute is based on a market price of \euro{23}. Two higher levels set at \euro{30} and \euro{35}, and a lower level at \euro{15}. The increase of 30\% and 50\% are based on \citet{Menozzi2020}, which studied preferences of sea food with environmental labels in Europe\footnote{\citet{Menozzi2020}  include an increase of 30\% of the average price market for an environmental sea food label. The increase up to 50\% corresponds to the case of a product also produced locally.}.

\begin{figure}[!h]
\centering
% \begin{minipage}{\textwidth}
  \includegraphics[width=0.9\textwidth]{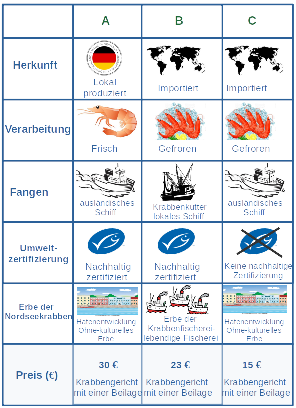}
  \caption[]{Choice Card for the participant in the experiment} 
  \label{fig:choiceCard}
 % \end{minipage}
\end{figure}

The participants were informed that the baseline (option C) was a hypothetical scenario  portraying the least favoured combinations of attributes of a shrimp dish. In comparison to the baseline, I expect that the alternative levels exhibit positive and significant values. The design consists of 16 choice cards divided into 4 blocks (each with four cards). Participants were assigned randomly to one of these blocks, enabling each respondent to make four choices, each one among three alternatives (A, B or C) (Table \ref{tab:attributesLevels}). The design of the choice cards was performed using R software following \citet{Aizaki2008} ensuring a D-efficient design and non-dominated solutions with a minimal degree of correlation among attributes in the design (Figure \ref{fig:choiceCard}).

\begin{table}[!htbp]
\centering
\resizebox{14cm}{!}{
\begin{tabular}{@{\extracolsep{5pt}}p{2.3cm}p{10cm}p{5cm}} 
%\begin{tabular}{c|c|c|c|c|c}
\hline
\\[-1.8ex]
Attribute   & Description & Levels\\
\\[-1.8ex]
\hline
\\[-1.8ex]
Origin & Country where the shrimp is harvested & Locally produced or Imported\\
\\[-1.8ex]
Processing & State of the shrimp before cooking & Fresh or frozen\\
\\[-1.8ex]
Harvesting & The practice that relates catching with local vessels (\textit{krabenkutter}) operating inshore versus foreign vessels. & Local vessel or foreign vessel \\
\\[-1.8ex]
Environmental certification & Shrimp harvested with a environmental conditions and received a eco-label & Sustainable certified or no-sustainable\\
\\[-1.8ex]
Heritage & Visual attribute of inshore fishing that relates to the possibility to enjoy cultural aspects such as access to visibly active shrimp fishing vessels operating at docks compared to a situation where access to the waterfront is restricted to areas redeveloped for residential and non-fishing commercial uses. & Waterfront development or fishing heritage\\
\\[-1.8ex]
Price & The willingness to pay for a shrimp meal with a side at a restaurant. & \euro{15}, \euro{23}, \euro{30}, \euro{35} \\
\\[-1.8ex]
\hline
\end{tabular}
}
\caption{Levels of attributes for each choice cards}
\label{tab:attributesLevels}
\end{table}

%CHOICE CARD

\subsection{Econometric Framework}

I analysed the DCE using the random utility theory. The conceptual theory is based on a scenario  in which a person or decision-maker $n$ faces a choice among $J$ alternatives. The decision-maker $n$ derives a specific level of utility from each alternative $j$, denoted $U_{nj}$, $j=a,b,c$. The decision-maker chooses the alternative that provides the greatest utility, choosing an alternative $i$ if and only if $U_{ni} > U_{nj} \forall j \neq j$.

This utility is known to the decision-maker but not to the researcher; therefore, it comprises an observed component $V_{nj}$ and a random stochastic component $\epsilon_{nj}$. Hence the utility can be represented as $U_{nj} = V_{nj} + \epsilon_{nj}$. Often, the observed part of the utility is specified in linear parameters, where $x_{nj}$ is a vector of variables (attributes) that relate to alternative $j$ as faced by the decision maker $n$, and $\beta$ is a vector of coefficients for these variables. The error term, $\epsilon_{ni}$, is assumed to be independent and identically distributed (IID) extreme value. The utility function is described in equation (\ref{eq:utility}), where each attribute represents a dummy variable indicating its presence or absence in the alternative $j$, expect for \textit{Price}, which is described as a categorical variable.

\begin{equation}
\begin{split}
U_{njt} = \beta_0 + \beta_1 +  \beta_{n2} Price_{njt} + \beta_{n3}Origin_{njt} + \beta_{n4}Processing_{njt} + \\
 \beta_{n5}Harvesting_{njt} + \beta_{n6}Certification_{njt} +  \beta_{n7}Heritage_{njt} + \epsilon_{njt}  
\end{split}
\label{eq:utility}
\end{equation}  

To estimate the probability of choosing the alternative $i$, I used the Random Parameters Mixed Logit Model (RPL). This model aims to capture preference heterogeneity and allows random taste variation among individuals. The probability of individual $n$ choosing the alternative $j$ is derived conditional to the density $f$ of the coefficients $\beta$, with parameters $\theta$ referring to the mean and variance of $\beta$ ($f(\beta \mid \theta)$). In this context, the distribution of $\beta$ is estimated using its mean and standard deviation (Equation \ref{eq: rpmlProbability}). 

\begin{equation}
P_{ni} = \int \left( L_{ni}(\beta) \right) f(\beta \mid \theta) d \beta
\label{eq: rpmlProbability}
\end{equation}

$L_{ni}(\beta) = \frac{e^{\beta' x_{ni}}}{\sum_j e^{\beta' x_{nj}}}$  represents the probability of choosing the alternative $i$ for a given $\beta$. There are no closed form solutions for equation (\ref{eq: rpmlProbability}), hence the probabilities are approximated through simulation techniques for any given value of $\theta$.   Conditional on $\beta$, the probability that the person $n$ makes this sequence of choices is the product of logit formulas:

\begin{equation}
\textbf{L}_{n\textbf{i}}(\beta) = \Pi_{t=1}^{T}  \left[ \frac{e^{\beta_{n} x_{ni_t t}}}{\sum_{j}e^{\beta_n x_{njt}}} \right]
\end{equation} 

The unconditional probability is the integral of this product over all values of $\beta$: $P_{n\textbf{i}} = \int \textbf{L}_{n\textbf{i}}(\beta) f(\beta) d\beta$. Where $\textbf{i}$ represents a sequence of alternatives, one for each choice situation $t$ $(\textbf{i}= \lbrace i_1, .., i_T \rbrace ) $.  The parameters of the model are estimated using a maximum likelihood estimation technique. The distribution density of the random parameters was set to normal and the number of draws was set to 100. I implemented this model in R software following \citep{Croissant2020}.

The marginal Willingness To Pay (WTP) for each attribute is determined by the change in price associated with a unit increase in that attribute. WTP represents the value that the average respondent $n$ is willing to pay for an increase of one unit in the given attribute. Recall that the coefficient of the attribute corresponds to each element in the vector $\beta_n$.

\begin{equation}
WTP= \beta_{attribute} / \beta_{price}
\end{equation}            

Given the potential correlation among the campaign visibility and the probability of choosing the alternative $j$, I used the expansion of coefficients method to identify heterogeneous preferences following \citet{Paez2022}. The model aims to find significant correlations among the campaign presence in the town and the respondents' decisions, an interaction term was added to equation (\ref{eq:utility}) (See Equation \ref{eq:utilityHet1}). 

\begin{equation}
\begin{split}
U_{njt} = \beta_0 + \beta_1 +  \beta_{n2} Price_{njt} + ,...., +  \beta_{n7}Heritage_{njt} +\\ \beta_{8}Heritage_{njt}*Campaign_{nt} + \epsilon_{njt}  
\end{split}
\label{eq:utilityHet1}
\end{equation}

\section{Results}
\subsection{Descriptive Statistics}

The participants in the survey showed a diverse range of socio-demographic characteristics (Table \ref{tab:descriptiveStatistics}). Among the tourists, 80\% belonged to B\"{u}sum and Greetsiel, and among the locals 55\% belonged to Cuxhaven. On average, less than 13\% of respondents reported a connection to the fisheries sector, whether through relatives or acquaintances; this connection was more prevalent in Ditzum and Cuxhaven.	
Sixty-five participants provide no information about their income, but among those who did, an average of 49\% earned more than \euro{40.000} per year. In the hypothetical scenario, with no more shrimp fishing and neither vessels on the harbour, over 70\% of participants reported to \textit{come back} to these towns, with Ditzum showing the lowest and Cuxhaven the highest proportion. Most respondents came from the states of Niedersachsen and North Rhine-Westphalia, accounting for 38\% of the sample (See section \ref{fig:mapTorisim}).

\begin{table}[!htbp]
\centering
\caption{Descriptive statistics of the sample}
\begin{tabular}{@{\extracolsep{5pt}}p{2.6cm}ccccc} 
\hline
\\[-1.8ex]
   & All & Busum & Ditzum & Greetsiel & Cuxhaven\\
\\[-1.8ex]
\hline
\\[-1.8ex]
Age & 54.414 & 52.752 & 58.852 & 54.971 & 53.301\\
\\[-1.8ex]
Female \% & 0.550 & 0.523 & 0.419 & 0.600 & 0.594\\
\\[-1.8ex]
Tourist \% & 0.765 & 0.872 & 0.774 & 0.857 & 0.602\\
\\[-1.8ex]
Education & 13.459 & 12.596 & 14.918 & 13.942 & 13.120\\
\\[-1.8ex]
Income & 3.603 & 3.876 & 3.784 & 3.670 & 3.692\\
\\[-1.8ex]
Connection \% & 0.125 & 0.083 & 0.194 & 0.076 & 0.165\\
\\[-1.8ex]
Come Back \% & 0.778 & 0.844 & 0.452 & 0.724 & 0.917\\
\\[-1.8ex]
\hline
\\[-1.8ex]
Sample Size & 409 & 109 & 62 & 105 & 133\\
\hline
\end{tabular}
\caption[Descriptive statistics of the sample]{Descriptive statistics of the sample per town. \textit{Age }displays the average values per town. \textit{Female} indicates the percentage of women, while \textit{Tourist } represents the percentage of people who were tourists. The variable \textit{income} is categorical,ranging from 1 to 6,  with each category increasing by \euro{10,000}, starting at \euro{20,000} as annual income before taxes. \textit{Education} represents the number of years of education. \textit{Connection} illustrates the percentage of people who claim to have a connection with fisheries, either through relatives or work-related contacts in the fishing industry. \textit{Connection} takes the value of 1 if the person had some connection with fishing and zero otherwise. \textit{Come back} reflects the percentage of people who intend to return to the town, even if there is no longer any shrimp fishing cultural heritage available for tourists.}
\label{tab:descriptiveStatistics}
\end{table}

Participants' reasons for living in or visiting these towns varied across towns (Table \ref{tab:PurposevisitDescriptions}). Nature emerged as the most important reason, encompassing activities such as hiking, biking, walking, and sightseeing. Cuxhaven and B\"{u}sum recorded the highest averages in the Nature item, with a statistically significant difference compared to Ditzum and Greetsiel. Water activities followed a similar trend, largely due to the beach proximity of Cuxhaven and B\"{u}sum, which serves as a major attraction for tourists. Additionally, Work-related reasons showed significant differences among towns, with Cuxhaven having the highest proportion of participants who live or visit for work purposes.

%[!htbp]
\begin{table}[ht!]
\centering
\caption{Purpose of visit per town. Average from the scale 1 (not important) to 5 (very important). }
\resizebox{12cm}{!}{
\begin{tabular}{cccccc}
\hline
\\[-1.8ex]
Visit & Ditzum & Greetsiel & Cuxhaven & Büsum &  P value\\
\\[-1.8ex]
\hline
\\[-1.8ex]
Work & 1.45 & 1.34 & 1.96 & 1.34  & 0.029\\
\\[-1.8ex]
Family & 2.93 & 1.56 & 2.47 & 1.68& 0.756\\
\\[-1.8ex]
Fishery & 3.53 & 3.70 & 3.51 & 3.60  & 0.532\\
\\[-1.8ex]
Food & 3.63 & 3.37 & 3.38 & 3.39  & 0.608\\
\\[-1.8ex]
Nature & 3.95 & 4.81 & 4.69 & 4.87  & 0.005\\
\\[-1.8ex]
Water Activities & 2.20 & 2.29 & 3.02 & 3.35  & 0.000\\
\\[-1.8ex]
\hline
\end{tabular}
}
\caption[Purpose of visit per town]{The rows show the purpose of the visit for each town. The numbers indicate the average ratings on a scale from 1 (not important) to 5 (very important). The last column shows the P-Value regarding significant differences among towns with a visible campaign (Greetsiel and Ditzum) and non-visible campaign (Büsum and Cuxhaven).}
\label{tab:PurposevisitDescriptions}
\end{table}

\newpage

\subsection{Cultural heritage preferences}
\label{sec:culHeritagePref}

Table \ref{tab:heritageAttitScale} presents the results of the attitudinal scale towards the shrimp cultural heritage across five factors. Factor one,  intercommunity and inter-generational linkages, showed significantly higher values in towns with a visible campaign. Factor two, recognition of diverse cultural values, also had higher values in these towns, although only three of the four items were statistically significant. Factors three and four reported no statistical difference among towns with a visible and non-visible campaign, however B\"{u}sum had the highest average for preserving cultural traditions in factor four. Lastly, factor five showed significantly higher values for towns with a visible campaign.

%!htbp
\begin{table}[ht!]\centering
\caption{Average of the survey regarding the attitudinal scale towards the shrimp cultural heritage}
\resizebox{13cm}{!}{ 
%\resizebox{9cm\textwidth}{!}{
\begin{tabular}{>{\raggedright\arraybackslash}cp{10cm}ccccccc}
\hline
\\[-1.8ex]
\# & item & All & Ditzum & Greetsiel & Busum & Cuxhaven & F-Test-All & F-Test-Campaign\\
\\[-1.8ex]
\hline
1 & It is important to maintain maritime cultural heritage & 4.69 & 4.77 & 4.72 & 4.71 & 4.62 & 1.454 & 2.462\\
\\[-1.8ex]
& \textbf{Factor 1: Intercommunity and intergenerational linkages} &   &      &      &      &       & &   \\
\\[-1.8ex]
2 & I am glad because shrimp cultural heritage is available to me & 4.31 & 4.56 & 4.45 & 4.27 & 4.11 & 5.417$^{**}$ & 13.456$^{***}$\\
\\[-1.8ex]
3 & We do need to conserve the shrimp cultural heritage for future generations & 4.50 & 4.73 & 4.59 & 4.45 & 4.37 & 3.914$^{**}$ & 9.772$^{**}$\\
\\[-1.8ex]
4 & The cultural values present in the Shrimp fishery heritage of our forefathers are important to me. & 4.00 & 4.15 & 4.07 & 3.99 & 3.87 & 1.424 & 3.105$^{*}$\\
\\[-1.8ex]
 & \textbf{Factor 2: Recognition of diverse cultural values} &   &      &      &      &       &  \\
\\[-1.8ex] 
5 & Shrimp fishing  heritage helps me to identify myself & 2.44 & 3.02 & 2.50 & 2.27 & 2.26 & 5.242$^{**}$ & 9.755$^{**}$\\
\\[-1.8ex]
6 & We need to take care about shrimp cultural heritage & 4.25 & 4.53 & 4.32 & 4.28 & 4.04 & 4.901$^{**}$ & 8.169$^{**}$\\
\\[-1.8ex]
7 & We have the right to destroy the shrimp cultural heritage to suit our needs & 1.33 & 1.48 & 1.36 & 1.17 & 1.35 & 1.970 & 2.454\\
\\[-1.8ex]
8 & I recognize the existence of shrimp cultural heritage in this town (in food, vessels, architecture) & 3.98 & 4.15 & 4.28 & 4.07 & 3.59 & 10.176$^{***}$ & 16.139$^{***}$\\
\\[-1.8ex]
& \textbf{Factor 3: Awareness of cultural loss} &   &      &      &      &       &  &\\
\\[-1.8ex]
9 & If things continue on their present course, we will soon experience a major loss in shrimp fishery cultural heritage. & 3.97 & 3.84 & 3.97 & 4.09 & 3.94 & 1.011 & 0.793\\
\\[-1.8ex]
10 & The shrimp fishery  cultural heritage is disappearing & 3.25 & 3.42 & 3.17 & 2.99 & 3.45 & 3.347$^{*}$ & 0.052\\
\\[-1.8ex]
 & \textbf{Factor 4: Preservation of traditions and customs} &   &      &      &      &       &  &\\
\\[-1.8ex]
11 & I want to know the traditions of our grand parents who practice the shrimp fishery & 3.30 & 3.37 & 3.12 & 3.47 & 3.28 & 0.741 & 0.725\\
\\[-1.8ex]
  & \textbf{Factor 5: Economic, environmental and social dimensions} &   &      &      &      &       &  &\\
\\[-1.8ex]
12 & Local shrimp fishing it is economically important for the fishers & 4.45 & 4.63 & 4.63 & 4.44 & 4.24 & 7.614$^{***}$ & 17.829$^{***}$\\
\\[-1.8ex]
13 & Local shrimp fishing is a tourist attraction, i.e. instrument for local economic development & 3.95 & 4.37 & 4.20 & 4.14 & 3.41 & 19.794$^{***}$ & 25.049$^{***}$\\
\\[-1.8ex]
14 & Local shrimp fishing reminds me about the connection with the sea/environment & 4.00 & 4.52 & 4.06 & 3.97 & 3.73 & 8.347$^{***}$ & 13.613$^{***}$\\
\\[-1.8ex]
15 & Local shrimp fishing influences the character of the place through buildings, symbols, traditions, etc. It is part of the cultural heritage. & 4.12 & 4.43 & 4.39 & 4.09 & 3.79 & 11.443$^{***}$ & 27.259$^{***}$\\
\\[-1.8ex]
16 & Local shrimp fishing is important to be done sustainably respecting living and non-living resources in the sea. & 4.76 & 4.84 & 4.82 & 4.80 & 4.63 & 3.303$^{*}$ & 4.500$^{*}$\\
\hline
 &\textit{Note:}  & \multicolumn{7}{r}{$^{*}$p$<$0.1; $^{**}$p$<$0.05; $^{***}$p$<$0.01} \\ 

\end{tabular}
}
\caption[Average values of the cultural heritage survey]{Average values of the survey regarding the attitudinal scale towards the shrimp cultural heritage. The scale from one to five, where (1) means: ``I do not agree at all with the statement" and (5) means ``I agree totally with the statement". The column F-Test-All shows the  F-test for significant differences among towns. The last column (F-Test-Campaign) indicates significant differences regarding the presence of the campaign, high visible and low visible campaign  (Ditzum and Greetsiel vs B\"{u}sum and Cuxhaven).}
\label{tab:heritageAttitScale}
\end{table}

Among all items, the highest average score was regarding the affirmations: (1) “Local shrimp fishing is important to be done sustainably respecting living and non-living resources in the sea", (2) “It is important to maintain maritime cultural heritage", and (3) “We do need to conserve the shrimp cultural heritage for future generations". This result showed that respondents gave the highest importance to sustainability of the fishing practice, followed by maintaining this practice for future generations.

Based on the attitudinal scale results, I conducted a Hierarchical Cluster Analysis (HCA) to categorize individuals according to their preferences for shrimp cultural heritage, as outlined by  \citet{Choi2007}. I employed the Ward method that produces the highest agglomerative coefficient, resulting in a better fit than other methods \footnote{The  squared Euclidean distance served as the measurement of the distance matrix}. The clustering gap statistic indicated that the optimal number of clusters was five; however, the gap statistic was very similar for two or five clusters. For ease of analysis, I divided the sample in two clusters: Individuals with (a) High preferences (pro-heritage) and (b) Low preferences for shrimp cultural heritage (No pro-heritage).  

I performed a logistic regression to identify individual explanatory characteristics related with pro-heritage attitudes (Table \ref{tab:clusterResults}). The dependent variable is coded as one if the individual belongs to the cluster with a high preference for shrimp cultural heritage (pro-heritage) and zero otherwise. The results indicate a significant correlation between older individuals and those less educated with higher cultural preferences. The variables of age and years of education were negative and significantly correlated, on average  older people were less educated (correlation=-0.21, p-value = 1.433e-05). Additionally, individuals located in towns with a high visible campaign showed a positive significant correlation with a higher attitude towards shrimp cultural heritage, i.e, a higher probability of belonging to the pro-heritage group.

\begin{table}[ht!] 
\centering 
\resizebox{9cm}{!}{   
\begin{tabular}{@{\extracolsep{5pt}}lcc } 
\\[-1.8ex]\hline 
\hline \\[-1.8ex] 
 & \multicolumn{2}{c}{\textit{Dependent variable:}} \\ 
\cline{2-3} 
\\[-1.8ex] & \multicolumn{2}{c}{Preferences for cultural heritage (Pro-Heritage)} \\ 
\\[-1.8ex] & \multicolumn{1}{c}{(1)} & \multicolumn{1}{c}{(2)}\\ 
\hline \\[-1.8ex] 
 Female & -0.088 & -0.090 \\ 
  & (0.214) & (0.235) \\ 
  Age & 0.013$^{*}$ & 0.018$^{**}$ \\ 
  & (0.007) & (0.007) \\ 
  Education & -0.072$^{**}$ & -0.084$^{**}$ \\ 
  & (0.034) & (0.039) \\ 
  Income &  & -0.029 \\ 
  &  & (0.073) \\ 
  Campaign & 0.626$^{***}$ & 0.461$^{*}$ \\ 
  & (0.225) & (0.250) \\ 
  Fix Income & 0.385 & 0.292 \\ 
  & (0.298) & (0.321) \\ 
  Tourist & -0.247 & 0.018 \\ 
  & (0.254) & (0.276) \\ 
  Constant & 0.213 & 0.219 \\ 
  & (0.708) & (0.760) \\ 
 \hline \\[-1.8ex] 
Observations & \multicolumn{1}{c}{395} & \multicolumn{1}{c}{334} \\ 
Log Likelihood & \multicolumn{1}{c}{-257.973} & \multicolumn{1}{c}{-215.587} \\ 
Akaike Inf. Crit. & \multicolumn{1}{c}{529.945} & \multicolumn{1}{c}{447.174} \\ 
\hline 
\hline \\[-1.8ex] 
\textit{Note:}  & \multicolumn{2}{r}{$^{*}$p$<$0.1; $^{**}$p$<$0.05; $^{***}$p$<$0.01} \\ 
\end{tabular}
}
  \caption{Logit estimation with preferences for cultural heritage. \textit{Campaign} has the value of one if the participant was located in a town with a high-visibility campaign and zero otherwise. \textit{Fix Income} is one if the individual is either retired or employed and zero otherwise. } 
  \label{tab:clusterResults} 
\end{table}

\newpage

\subsection{Choice experiments results}

The results of the Random Parameters Mixed Logit (RPL) analysis indicate that the null hypothesis- that the coefficients are zero - is rejected, all coefficients are statistically significant at the 1\% level (Table \ref{tab:resultsMIXED}). The alternative specific constants for choices A and B indicate changes in utility relative to alternative C. The results indicate that, on average, respondents prefer choices A or B over alternative C. This implies that, people prefer a shrimp dish with at least one of the attributes rather than having none. The coefficient of the price is negative, as expected, indicating that an increase in price leads to lower utility and a decreased likelihood of purchase.

\begin{table}[ht]
\centering
\resizebox{12cm}{!}{  
\begin{tabular}{@{\extracolsep{5pt}}lcccc} 
\hline
\hline
\\[-1.8ex]
  & (1) & (2) & (3) & (4)\\
\\[-1.8ex]
\hline
\\[-1.8ex]
Choice A & 1.361$^{***}$ & 1.359$^{***}$ & 1.342$^{***}$ & 1.336$^{***}$\\
\\[-1.8ex]
Choice B & 1.580$^{***}$ & 1.582$^{***}$ & 1.539$^{***}$ & 1.536$^{***}$\\
\\[-1.8ex]
Price & -0.090$^{***}$ & -0.090$^{***}$ & -0.089$^{***}$ & -0.089$^{***}$\\
\\[-1.8ex]
\textbf{Origin} & 0.965$^{***}$ & 0.964$^{***}$ & 0.979$^{***}$ & 0.977$^{***}$\\
\\[-1.8ex]
Std Dev & 1.028$^{***}$ & 1.014$^{***}$ & 0.965$^{***}$ & 0.952$^{***}$\\
\\[-1.8ex]
\textbf{Processing} & 0.624$^{***}$ & 0.634$^{***}$ & 0.625$^{***}$ & 0.629$^{***}$\\
\\[-1.8ex]
Std Dev & 0.767$^{***}$ & 0.770$^{***}$ & 0.659$^{**}$ & 0.639$^{*}$\\
\\[-1.8ex]
\textbf{Harvesting} & 0.938$^{***}$ & 0.944$^{***}$ & 0.960$^{***}$ & 0.964$^{***}$\\
\\[-1.8ex]
Std Dev & 1.069$^{***}$ & 1.098$^{***}$ & 1.110$^{***}$ & 1.125$^{***}$\\
\\[-1.8ex]
\textbf{Certification} & 1.215$^{***}$ & 1.221$^{***}$ & 1.183$^{***}$ & 1.180$^{***}$\\
\\[-1.8ex]
Std Dev & 1.150$^{***}$ & 1.165$^{***}$ & 1.209$^{***}$ & 1.207$^{***}$\\
\\[-1.8ex]
\textbf{Heritage} & 0.454$^{***}$ & 0.221$^{}$ & 0.220$^{}$ & 0.062$^{}$\\
\\[-1.8ex]
Std Dev & 0.829$^{***}$ & 0.785$^{***}$ & 0.874$^{***}$ & 0.843$^{***}$\\
\\[-1.8ex]
\textbf{Heritage*Campaign} &      & 0.567$^{***}$ &  & 0.465$^{**}$\\
\\[-1.8ex]
\textbf{Heritage*Pro-Hert} &       &               & 0.414$^{**}$ & 0.355$^{*}$\\
\\[-1.8ex]
\hline
\\[-1.8ex]
AIC & 2758.700 & 2752.665 & 2679.953 & 2677.010\\
\\[-1.8ex]
BIC & 2828.716 & 2828.066 & 2754.950 & 2757.364\\
\\[-1.8ex]
Log-likelihod & -1366.350 & -1362.332 & -1325.976 & -1323.505\\
\\[-1.8ex]
\hline
\textit{Note:}  & \multicolumn{4}{r}{$^{*}$p$<$0.1; $^{**}$p$<$0.05; $^{***}$p$<$0.01} \\ 
\end{tabular}
}
\caption{Random Parameters Mixed logit Model estimates. Column (1) shows results from the specification (\ref{eq:utility}), column (2) includes an interaction term for people located in towns with a high visible campaign (Equation \ref{eq:utilityHet1}), column (3) shows the specification with an interaction term accounting for people with pro-heritage attitudes and column (4) includes both interactions campaign and pro-heritage.}   
\label{tab:resultsMIXED}
\end{table}

The results of the RPL model, as outlined in Equation (\ref{eq:utility}), are presented in table \ref{tab:resultsMIXED} (column (1)). The means and standard deviations per attribute provide the share of respondents with a positive purchasing probability. Overall, 83\% of the respondents had a positive likelihood for choosing a shrimp dish produced locally (origin), 77\% with a fresh attribute (processing), 80\% with a local vessel attribute (harvesting), 86\% with a certification attribute (certification), and 68\% with a heritage attribute (heritage). This implies that consumers' utility increases when choosing a shrimp dish with any of these attributes.

Table \ref{tab:resultsMIXED} (Column (2)) presents the heterogeneity estimates  based on Equation (\ref{eq:utilityHet1}). Participants in towns with a high-visibility campaign significantly correlate with a higher purchasing probability of a heritage attribute than those with a low-visibility campaign (\textit{Heritage*campaign}). Similar result holds for specification in column (3), which includes an interaction between the heritage attribute and the pro-heritage attitudes (\textit{Heritage*Pro-Hert}). As expected, pro-heritage individuals are more likely to purchase a heritage attribute than those with low pro-heritage attitudes. Column (4) includes two interactions, with the \textit{campaign} and \textit{pro-heritage} variables, both which are positive and significant. Given that the campaign variable is likely endogenous with respect to town characteristics, these results highlight a significant correlation between individuals in towns with high campaign visibility and their purchasing probabilities.

\subsubsection*{Willingness to Pay (WTP) for each attribute}

The WTP estimates for the RPL are presented in Table \ref{tab:WTP-Heterogeneity} where WTP (1) and WTP (2) show the results derived from specifications in Table \ref{tab:resultsMIXED} (1) and (4) respectively. The highest WTP is for a shrimp dish with a certification attribute, followed by local shrimp (origin=1) harvested by a local vessel (Harvesting=1). The heritage attribute ranks lowest in WTP, i.e, consumers are willing to pay in average \euro{5} more for a shrimp dish with a heritage attribute. The results reveal high standard deviations in WTP for each attribute, indicating considerable variability that may be linked to campaign's visibility or pro-heritage attitudes.

\begin{table}[ht]
\centering
\resizebox{11cm}{!}{  
\begin{tabular}{@{\extracolsep{5pt}}lcccc} 
\hline
\hline
\\[-1.8ex]
  & WTP (1) & SD & WTP (2) & SD\\
\\[-1.8ex]
\hline
\\[-1.8ex]
Origin & 10.683$^{***}$ & 11.378$^{***}$ & 11.026$^{***}$ & 13.324$^{***}$\\
\\[-1.8ex]
				  & (1.611) & (2.310) & (1.700) & (2.212)\\
\\[-1.8ex]
Processing & 6.902$^{***}$ & 8.483$^{***}$ & 7.104$^{***}$ & 0.698$^{*}$\\
\\[-1.8ex]
					& (1.455) & (2.912) & (1.511) & (1.981)\\
\\[-1.8ex]
Harvesting & 10.375$^{***}$ & 11.830$^{***}$ & 10.878$^{***}$ & 10.748$^{***}$\\
\\[-1.8ex]
 					& (1.523) & (2.353) & (1.641) & (2.389)\\
\\[-1.8ex]
Certification & 13.450$^{***}$ & 12.727$^{***}$ & 13.324$^{***}$ & 7.214$^{***}$\\
\\[-1.8ex]
				& (2.126) & (2.625) & (2.212) & (3.840)\\
\\[-1.8ex]
Heritage & 5.026$^{***}$ & 9.170$^{***}$ & 0.698$^{}$ & 12.700$^{***}$\\
\\[-1.8ex]
					& (1.165) & (2.556) & (1.981) & (2.455)\\
\\[-1.8ex]
Heritage*Pro$-$Hert &  &  & 4.009$^{*}$ & \\
\\[-1.8ex]
					&  &  & (2.374) & \\
\\[-1.8ex]
Heritage*Campaign &  &  & 5.243$^{**}$ & \\
\\[-1.8ex]
					&  &  & (2.430) & \\
\\[-1.8ex]
\hline
\textit{Note:}  & \multicolumn{4}{r}{$^{*}$p$<$0.1; $^{**}$p$<$0.05; $^{***}$p$<$0.01;
Standard errors in parenthesis} \\ 
\end{tabular}
}
\caption{Mean and standard deviation of WTP estimates of the RPL model (Unit \euro{}). Standard errors in parenthesis. }
\label{tab:WTP-Heterogeneity}
\end{table}

\begin{figure} [ht!]
\begin{center}
  \includegraphics[width=1\textwidth]{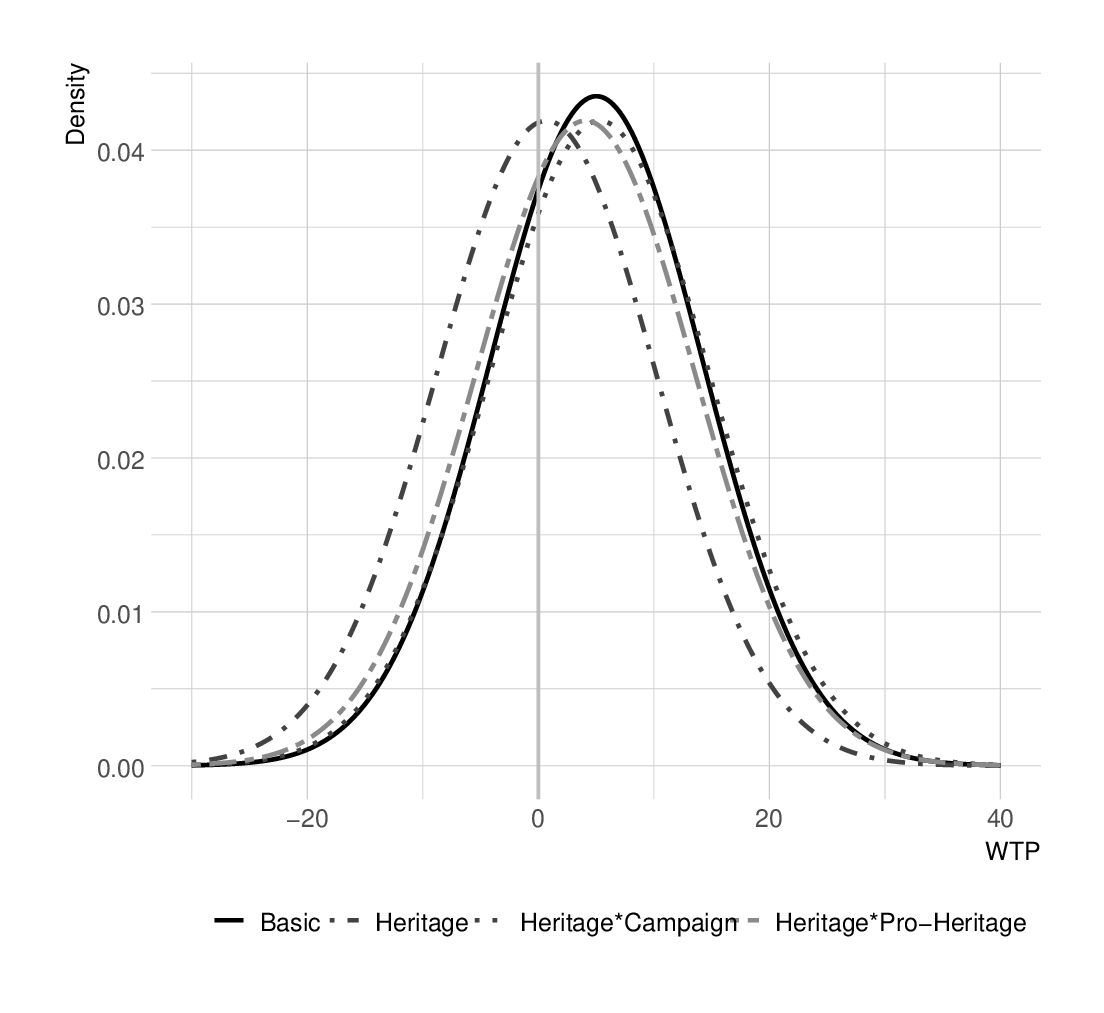}
  \caption{Kernel density functions for individual WTP regarding the heritage attribute derived from table \ref{tab:WTP-Heterogeneity}. The \textit{basic} line shows the WTP distribution for specification WTP (1). The remaining lines show the WTP densities for the second specification (WTP (2)). \textit{Heritage} corresponds to people with low pro-heritage attitudes located in low visible campaign towns, \textit{Heritage*Campaign} for people located in high visible campaign towns, and  \textit{Heritage*Pro-Heritage} for those with pro-heritage attitudes.} 
\label{fig:densitiesCluster}
\end{center}
\end{figure}

Table \ref{tab:WTP-Heterogeneity}, in column WTP (2), includes interaction terms to account for heterogeneity. People with high pro-heritage attitudes are willing to pay \euro{4} for the heritage attribute, while those located in towns with high visible campaign \euro{5}. Figure (\ref{fig:densitiesCluster}) shows the densities distribution of WTP for the heritage attribute. The basic line indicated that 68\% of people are willing to pay a positive amount for the heritage attribute (density from Equation \ref{eq:utility}). However, those with low pro-heritage attitudes and located in low visibility campaign towns are willing to pay \euro{0.69} on average.

\begin{table}[!ht]
\centering
\resizebox{12cm}{!}{  
\begin{tabular}{@{\extracolsep{5pt}}lcccccc} 
\hline
\hline
\\[-1.8ex]  
& Pro-heritage & No pro-heritage & p-Value & Campaign & No Campaign & p-Value\\
\\[-1.8ex]
\hline
\\[-1.8ex]
Female \% & 0.535 & 0.562 & 0.626 & 0.531 & 0.557 & 0.662\\
\\[-1.8ex]
Age & 56.722 & 51.385 & 0.001 & 56.208 & 53.262 & 0.069\\
\\[-1.8ex]
Education & 13.084 & 13.827 & 0.028 & 14.165 & 12.890 & 0.000\\
\\[-1.8ex]
Income & 3.572 & 3.688 & 0.548 & 3.707 & 3.571 & 0.478\\
\\[-1.8ex]
Fix Income & 0.873 & 0.787 & 0.020 & 0.838 & 0.835 & 0.846\\
\\[-1.8ex]
Tourist \% & 0.768 & 0.769 & 0.969 & 0.831 & 0.726 & 0.011\\
\\[-1.8ex]
\hline
\hline
\end{tabular}
}
\caption[Descriptive statistics for differences between sub-groups]{Descriptive statistics for differences between sub-groups. The p-value show the statistics results of the t-test mean differences between sub-groups.}
\label{tab:heterogeityAnalysis}
\end{table}

I divided the sample into four sub-groups to gain a deeper understanding of the WTP: Pro-heritage, No pro-heritage, campaign, and no-campaign (Table \ref{tab:heterogeityAnalysis}). Respondents located in towns with a highly visible campaign (\textit{campaign}) and low visible campaign (\textit{no-campaign})  significantly different in age, education, and percentage of tourists. Ditzum had the highest average age, while Cuxhaven contributed to the notable difference in tourist percentages, with nearly 40\% of respondents being locals (See table \ref{tab:descriptiveStatistics}). Among respondents in campaign towns, 62\% had high pro-heritage attitudes, compared with 51\% in no-campaign towns. This align with the  significant correlation between the presence of the campaign and the high pro-heritage attitudes (Table \ref{tab:clusterResults}). 

For each sub-group, I performed the same analysis derived from Equation (\ref{eq:utility}) using the RPL model to derive the WTP (Table \ref{tab:resultsSubGroups-proHeritage}). Among all the sub-groups, the certification attribute demonstrated the highest WTP, except for the campaign group, where it ranked second. Respondents in highly visible campaign towns ranked highest WTP for all attributes, followed by those with pro-heritage attitudes. Participants had the lowest WTP for a heritage attribute across all sub-groups, except for the campaign group, where it ranked second to last. To summarize, people with high pro-heritage attitudes or located in camping towns are willing to pay more for any shrimp dish attribute than people with low heritage preferences and located in towns with low campaign visibility.

\section{Discussion}

Maritime cultural heritage is recently becoming more relevant, but its integration into political decisions is challenging, due to the difficulty of capturing its value \citep{Hansen2022,Delaney2024, Less2023}. This study addresses this challenge within the German context by assessing the economic value of the Maritime Cultural Heritage associated with the brown shrimp fishery through a Discrete Choice Experiment (DCE). The findings reveal that this cultural heritage increases the utility of consumers when purchasing a shrimp dish. Respondents are willing to pay, on average,  an additional  of \euro{5.0} for a shrimp dish with a heritage attribute. Consumers also increase their utility by choosing a shrimp dish with any of the evaluated attributes compared to those without. These results  are novel, as this is the first study in Germany evaluating the value of shrimp cultural heritage, one of the oldest fishing techniques in the North Sea.

The Willingness To Pay (WTP) for maritime cultural heritage is heterogeneous across individuals. As expected and consistent with \citet{Duran2015}, those with high pro-heritage attitudes are willing to pay more than those who place less value on this heritage. The results show a significant correlation between pro-heritage attitudes and the visibility of an awareness-raising fishing campaign conducted by local fishers during this study. Respondents from towns with a highly-visible exhibit stronger pro-heritage attitudes and show a significant correlation with the WTP for the heritage attribute compared to those from towns with a less visible campaign. This study, however, does not establish causality; the campaign is likely endogenous to towns with strong historical ties to shrimp heritage, and visitors to these areas may inherently have pro-heritage preferences. Given the presence of the fishing campaign in all four towns, it is impossible to discern the effect of the campaign on WTP. It is likely that results are influenced by the presence of the campaign, and further research is needed to evaluate the effect of the campaign on the WTP by comparing towns with and without such campaigns.

%2. ROLE OF THE CAMPAGN IN THE WTP AND IMPORTANCE OF CULTURAL HERITAGE
Campaigns are well-known to increase consumers' WTP. Companies often use marketing campaigns to promote products and attract more customers, inducing them to increase their WTP. Many campaigns leverage moral causes, such as ethical narratives related to fair trade, sustainable, local product or charity causes, which can impact purchasing probabilities \citep{Park2018}.  Additionally, non-profit organizations like UNESCO conduct campaigns to raise awareness about the world cultural heritage sites, thereby enhancing peoples' willingness to engage in conservation activities \citep{Jaafar2015}. The results of this study align with existing literature, as they reveal a significant correlation between responses in campaign sites and higher WTP. Pro-heritage attitudes likely serve as the mechanism through which the campaign could influence WTP, given the significant correlation between these attitudes and the visibility of the campaign.

%3. OTHER FACTORS AFFECTING CULTURAL HERITAGE
Preferences for maritime cultural heritage are also affected by other factors such as age and education. The results of the attitudinal scale show that older and less educated individuals are significantly correlated with higher preferences for shrimp cultural heritage. In the sample, older people are generally less educated than younger generations. The correlation among age and years of education is negative and significant (corr= - 0.21 p-value=0.00001). The mean age of the sample in these towns is 54 years with a median of 58 (SD=16). A possible explanation for older people to be more pro-heritage is that they may have more memories, through magazines or media, about this fishery, a plausible reason to visit these areas. These findings align with those of \citet{Martino2023}, whose study also revealed a higher preference for fishing cultural heritage among local and older individuals compared to younger ones.

Among the evaluated attributes, certification is the most valued attribute, with consumers willing to pay an average of \euro{13.4}, followed by a product produced locally (\euro{10.7}). This result aligns with \citet{Martino2023, Tanner2021}, which also identify these two attributes as the most important influencing the WTP within the fishery sector. \citet{Bronnmann2021} conducted a DCE on sea food in the Baltic Sea also found that the certification label in fish products has the highest WTP among attributes analysed. Similarly, \citet{Menozzi2020}  found that in Germany, environmental labels are relevant for for herring, seabass and seabream compared to other European countries. \citet{Zander2018} found that European consumers value locally produced fish and are willing to pay a price premium to offset higher production costs. These findings collectively highlight the importance of certification and local production in shaping consumer preferences and WTP for seafood products.

The results indicate that the WTP for the heritage attribute is the lowest among evaluated attributes; however, it does not undermine its importance. This intangible good is often overlooked in policy reports and public opinion, particularly regarding the impact of Offshore Wind Farms (OWF) and other marine renewable on European fisheries \citep{EuropeanParliamentWilson2020}. While these reports discuss the potential for coexistence between offshore renewables and fisheries, they fail to address the impacts on cultural heritage resulting from declining fishing activities. \citet{Stelzenmuller2021} mention that the effect of OWF could result in economic losses and socio-cultural impacts in fisheries, but they lack assessment regarding the extend of the loss of maritime cultural heritage. 

The tangible and intangible benefits of preserving maritime cultural heritage are increasingly recognized. The European Commission acknowledge the tangible benefits of this heritage in terms of maritime and coastal tourism for economic development \citep{Delaney2024}. This study indicates that fishing -active shrimp fishing vessels and historical architecture- ranks as the second most important reason for visiting Greetsiel, Cuxhaven, and B\"{u}sum (Table \ref{tab:PurposevisitDescriptions}), indicating the potential of maritime cultural heritage to attract tourism and generate revenue for the German coastal region \citep{NDR20232}. In contrast, the intangible benefits of this heritage often lack assessed economic value, resulting in their low priority in cultural, economic and maritime policies \citep{Hansen2022}.

There are examples of the inclusion of intangible maritime cultural heritage in  political actions. In China, the government adopted measures at a national and county  levels to conserve and develop this heritage, integrating it with modern education, experiences facilities, responsible tourism, and even including it into school curriculum  \citep{Xiao2023}. Similarly, in the study area the towns of Greetsiel and Büsum adopted strategies to develop this intangible heritage, such as periodic conferences on the history of the brown shrimp fishery and recreation of traditional fishing experiences to enhance cultural awareness. Although these activities are useful, they could play a more prominent role. Survey results show that 67\% of the people express interest in getting to know the traditions of grand parents who practice shrimp fishery and most of the people are aware of the importance of preserving this tradition \footnote{In table \ref{tab:heritageAttitScale} factor 1 in average is greater than 4, indicating that people are aware of the importance of preserving the shrimp maritime cultural heritage.}. Given the demand for experiences and knowledge related to these traditions, there is an opportunity for policymakers to expand educational programs and community engagement initiatives. Developing this intangible heritage is crucial, as it fosters connection and belonging to a place, allowing to preserve the intangible aspects of this fishing heritage \citep{Xiao2023}.

In Europe, specifically in Scotland, France and Denmark, maritime cultural heritage is included in regional maritime development plans; however, these plans often emphasize tangible over intangible aspects \citep{Hansen2022}. In contrast, the western Mediterranean fisheries exclude the maritime cultural heritage in fisheries management, but contemplate the implementation of strategies such as Participatory Action Research (PAR) as a way to develop integrated policies  \citet{Gomez2021}. \citet{Hansen2022} and \citet{Gomez2021} argue that the absence of valuation studies for maritime cultural heritage diminishes its influence in political decisions \citep{Hansen2022}.  The first step toward integrating maritime cultural heritage into policies is recognizing its value; further challenges arise when integrating the cultural heritage aspect into Marine Spatial Planning (MSP) and urban development \citep{Hansen2022}. By identifying the intangible economic value of the maritime cultural heritage, this paper contributes to the ongoing discussions regarding the cultural, economic and  maritime policies in Germany and across in Europe \citep{Stelzenmuller2022, Gee2019, Hansen2022, Delaney2024}.

The implementation of integrated policies arise the question of financing the preservation and development of maritime cultural heritage, because governments funds are usually limited \citep{Xiao2023, Hansen2022}. Results of this study indicate that towns evaluated could increase their revenue from tourism. The survey shows that 96\% of the participants consider that the shrimp maritime cultural heritage is valuable for the present and future generations\footnote{This is the percentage of people who average more than 3 in the Factor 1 of the attitudinal preferences survey, the average in this factor is above 4 (See Table \ref{tab:heritageAttitScale}.)}. However, only 68\% of them  are willing to pay a positive amount to preserve this heritage. Yet, this percentage could increase by implementing strategies aimed at developing maritime heritage, besides only preserving it, what in turn increases value within society. 

Despite the increasing awareness of maritime cultural heritage, research remain limited, and fishing is often viewed solely as an economic activity focused on profits and jobs  \citep{Khakzad2016}. However, also contributes to a sense of place, fosters tourism, and helps to preserve old traditions \citep{Urquhart2013}. A location devoid of intangible cultural heritage lacks intrinsic value, relying instead on economic metrics that fluctuate with crises or climate change. The intrinsic value stems from the human connection to it, which goes beyond time. Balancing ecological, economic and sociocultural policies can prevent tipping points that lead to species extinction, economic crises, and the loss of traditions, which may ultimately exist only in written records. These findings highlight the recognition and consideration of maritime cultural heritage in policy discussions and decision-making processes.

\section{Conclusion}

This study contributes to increase the socio-cultural knowledge of the brown shrimp fishery in Germany. The results show that, besides an economic benefit, this fishery has a maritime cultural heritage value, being one of the oldest fishing techniques in the North Sea. Both residents and non-residents value the cultural heritage and are willing to pay an average of \euro{5} to sustain it. The WTP is significantly correlated with pro-heritage attitudes and the presence of a fishing campaign during the study period. Factors correlated with pro-heritage attitudes are age, education and the visibility of the fishing campaign. Notably, consumers exhibit the highest Willingness To Pay (WTP) of \euro{13.4} for a sustainable certification attribute,  followed by \euro{10.7} for a shrimp produced locally, and \euro{10.4} for a shrimp harvested with a local vessel. These results align with existing literature, particularly in the European seafood market. 

By assessing preferences for attributes of a brown shrimp dish and identifying pro-cultural attitudes of residents and visitors, I incorporate both informational and behavioural components into a policy advice where maritime cultural heritage is considered. This approach is essential for any information-based policy that influences purchasing decisions. I also draw on possible strategies directed to preserve and develop this heritage based on preferences obtained in the survey. The results of this study shed light on the broader implications of fishing policies that extent beyond the fishing activity itself, providing policy makers with a more comprehensive understanding for informed decision making. Future research could explore cultural resilience of fishers and its impact on ecological and social systems. While \citet{Quaas2013b} already investigated the relationship between consumer preferences and ecological consequences, the cultural aspect, specifically within fisheries socio-ecological systems, remains underexplored.

\textbf{Funding Sources}: This work was supported by The German Federal Ministry of Education and Science as part of the SeaUseTip Project [grant number 01LC1825C, 2018].

\printbibliography

\newpage  
\appendix 
  
\section{APPENDIX}
\label{sec:appendix1}

\subsection{Awareness-raising campaign for shrimp cultural heritage}

%!htbp
\begin{figure}[!h]
\centering
% \begin{minipage}{\textwidth}
  \includegraphics[width=1\textwidth]{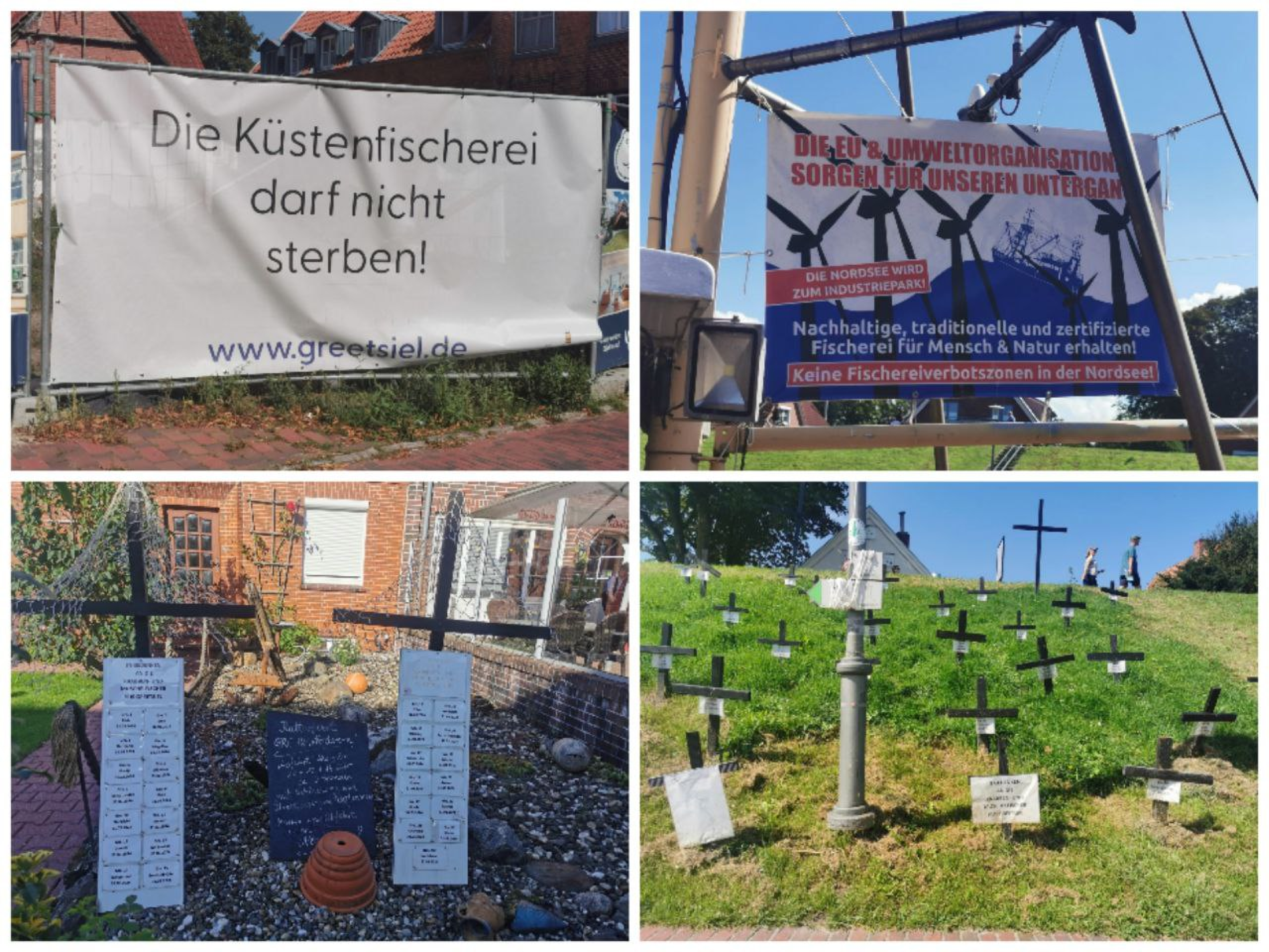}
  \caption[]{Campaign of the shrimp fisheries in Ditzum and Greetsiel} 
  \label{fig:photoCampaign}
  % \end{minipage}
\end{figure}

\begin{figure}[!htbp]
\centering
% \begin{minipage}{\textwidth}
  \includegraphics[width=1\textwidth]{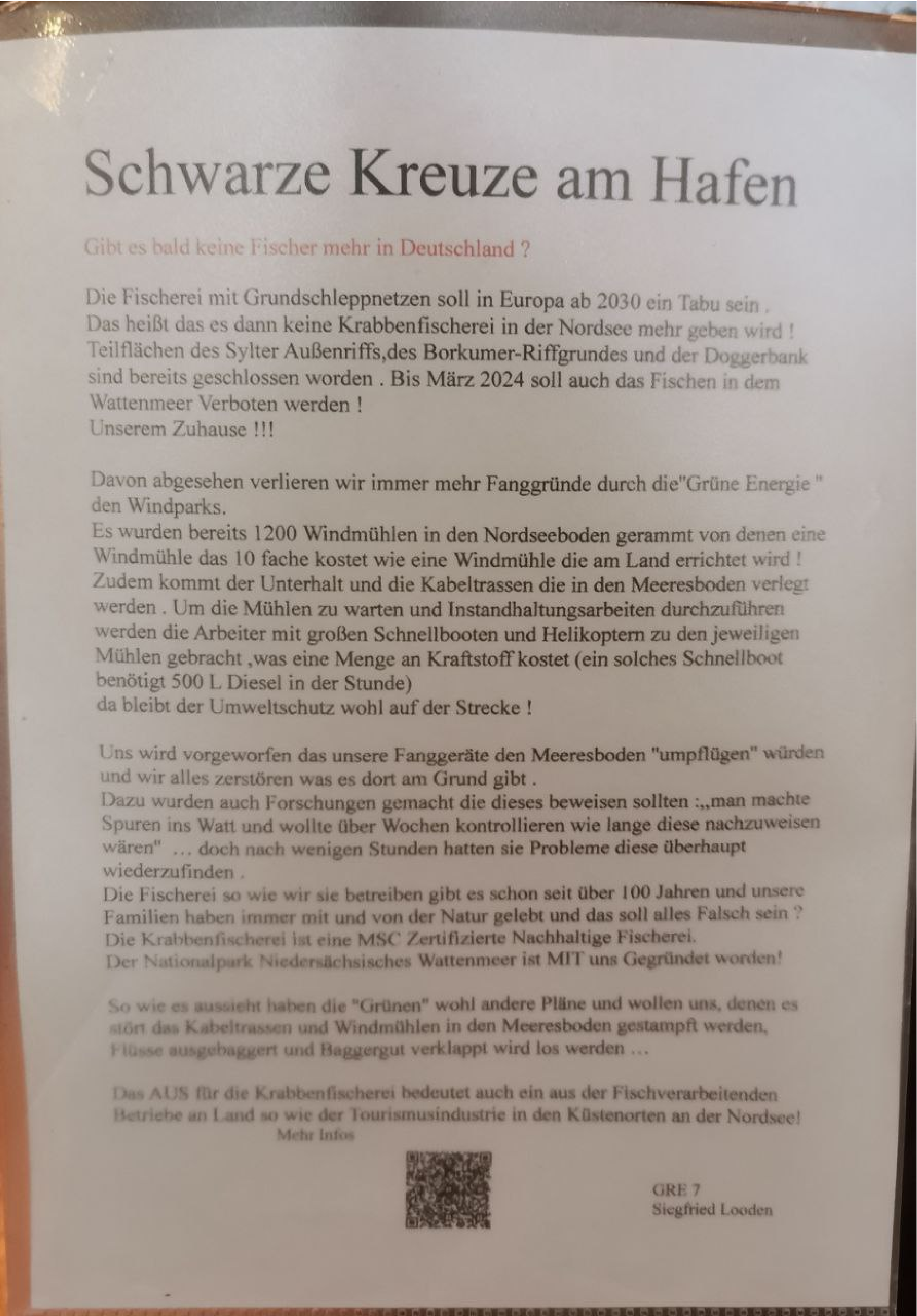}
  \caption[]{Letter of the campaign of the shrimp fisheries in Ditzum and Greetsiel} 
  \label{fig:photoLetter}
  % \end{minipage}
\end{figure}

\newpage
\subsection{Origin of the respondents in the sample}

\begin{figure}[!htbp]
\centering
% \begin{minipage}{\textwidth}
  \includegraphics[width=0.5\textwidth]{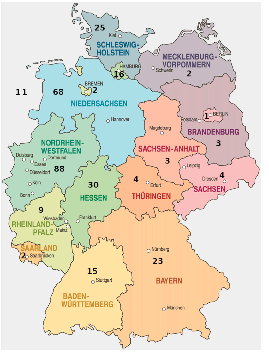}
  \caption[]{Quantity of tourist respondents per origin of federal state. The number outside of the map (11) corresponds to foreigners from Netherlands, Denmark, Belgium and Switzerland. } 
  \label{fig:mapTorisim}
 % \end{minipage}
\end{figure}

\newpage
\subsection{Survey Implemented in the study}

\begin{figure}[!htbp]
\centering
  \includegraphics[width=1\textwidth]{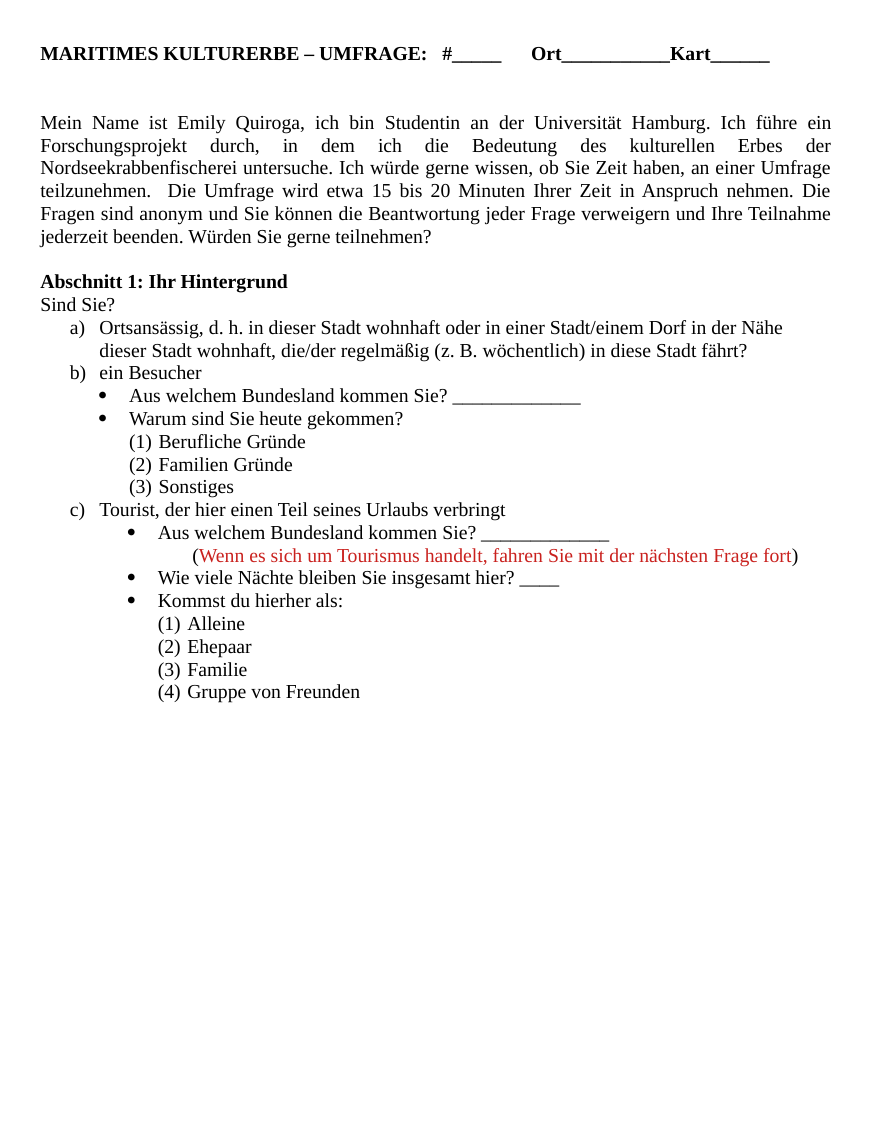}
\end{figure}

\begin{figure}[!htbp]
\centering
  \includegraphics[width=1\textwidth]{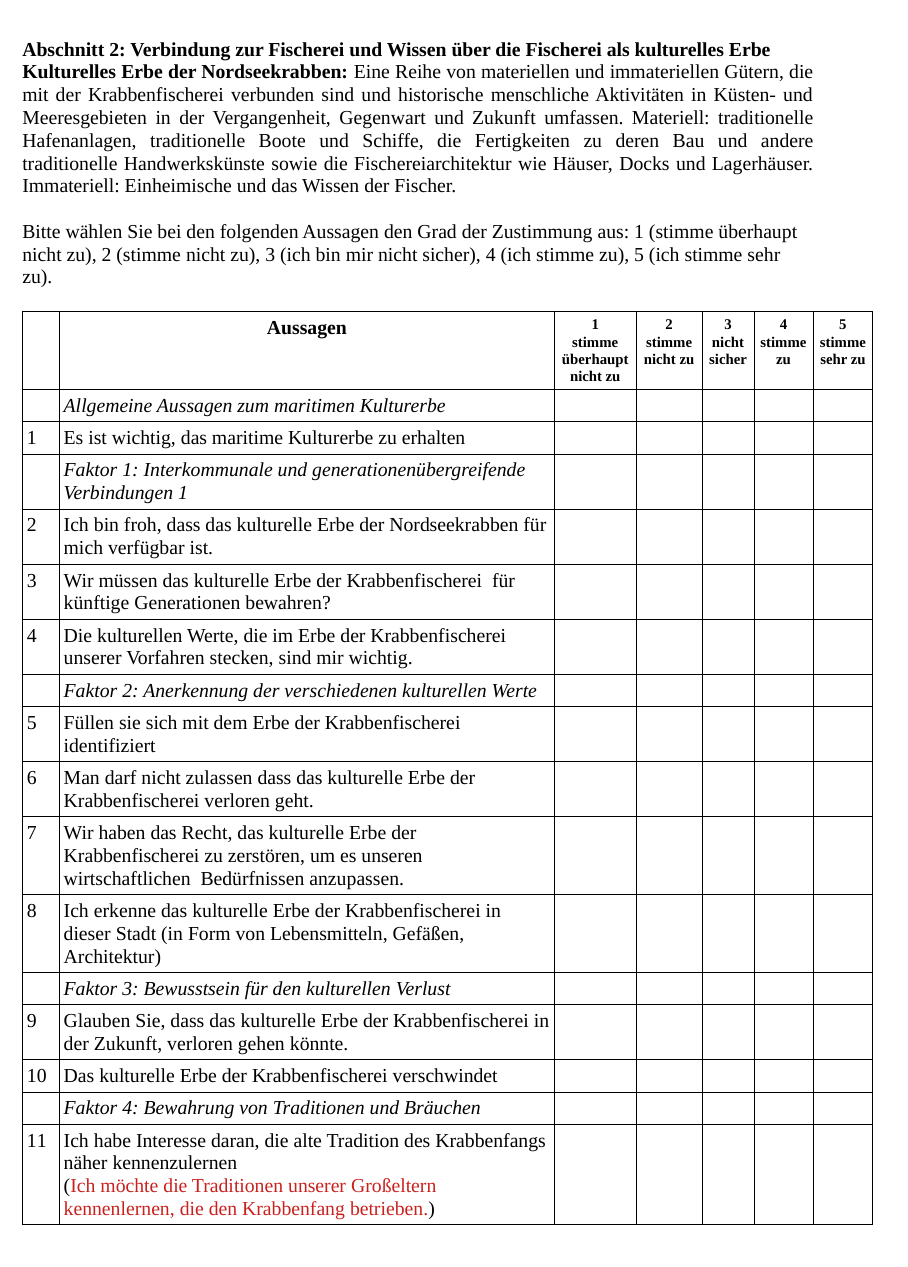}
\end{figure}

\begin{figure}[!htbp]
\centering
  \includegraphics[width=1\textwidth]{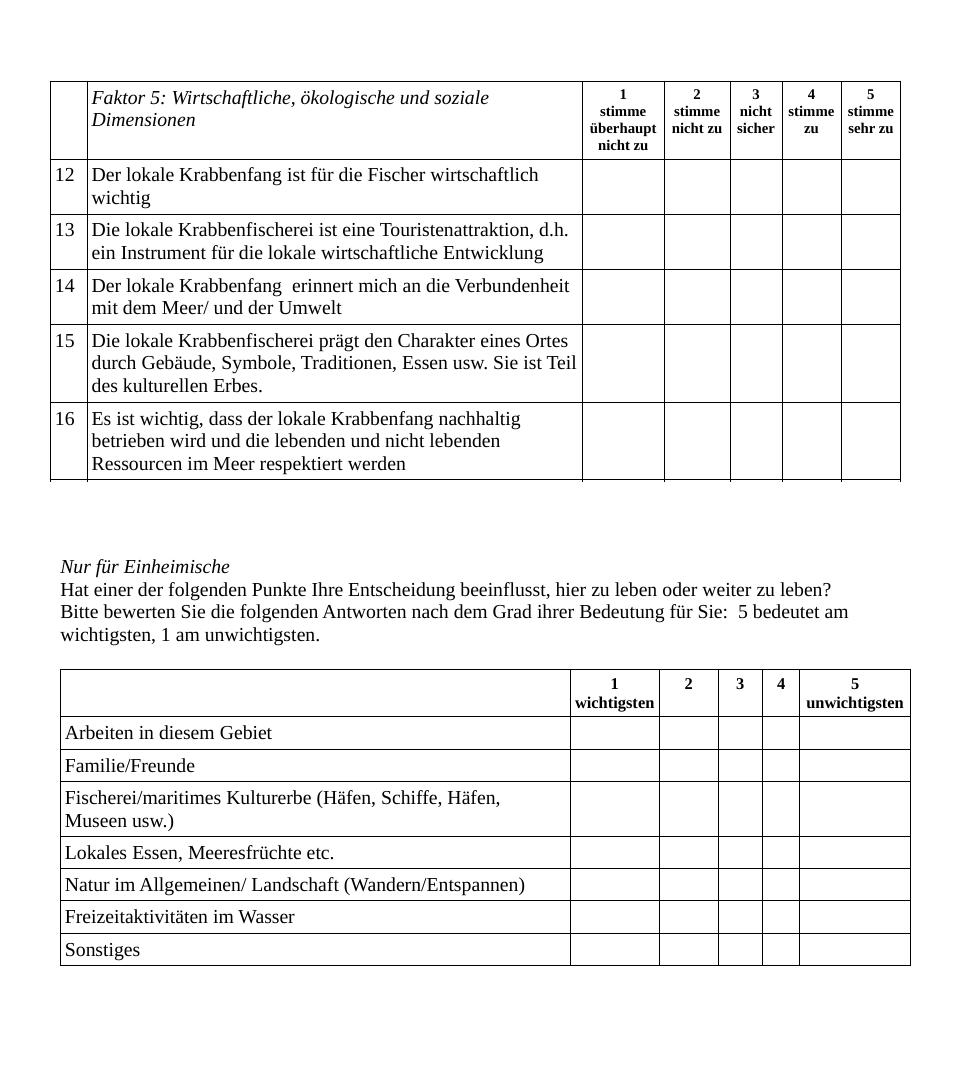}
\end{figure}

\begin{figure}[!htbp]
\centering
  \includegraphics[width=1\textwidth]{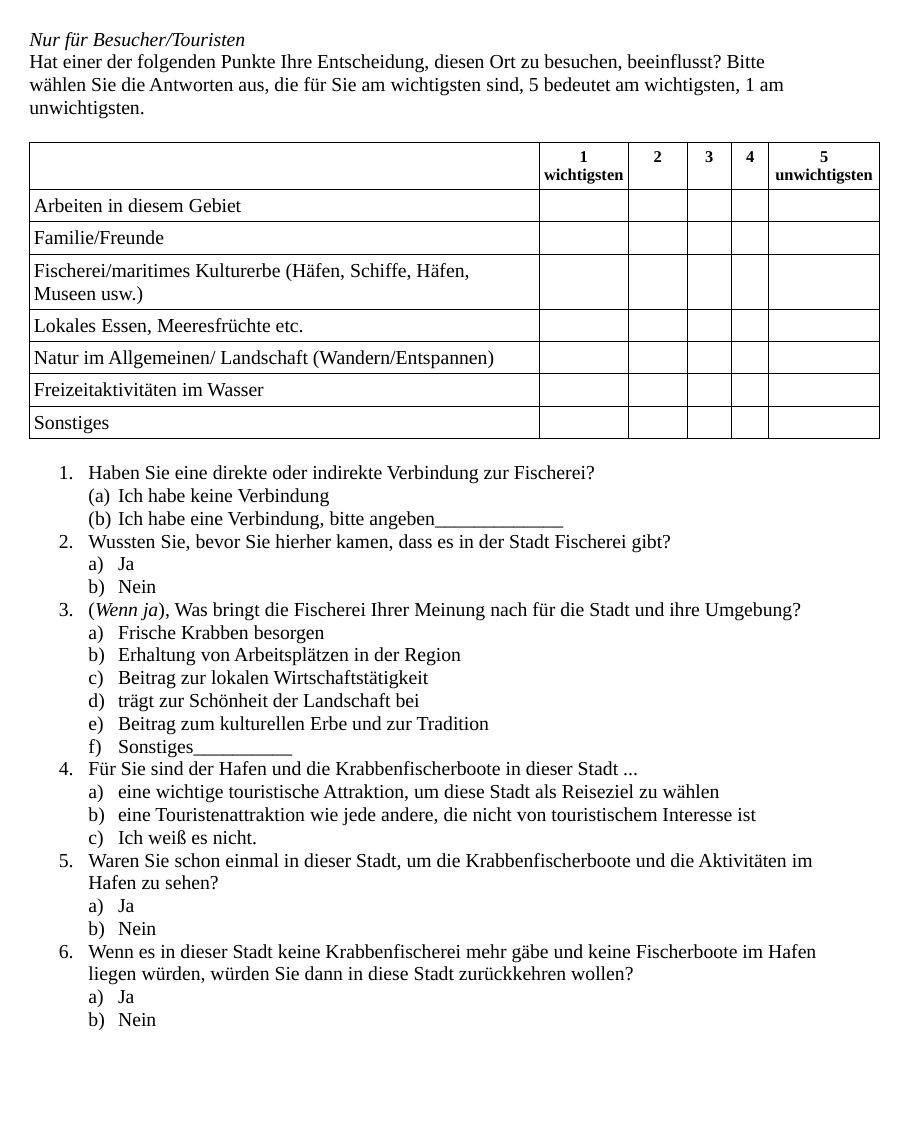}
\end{figure}

\begin{figure}[!htbp]
\centering
  \includegraphics[width=1\textwidth]{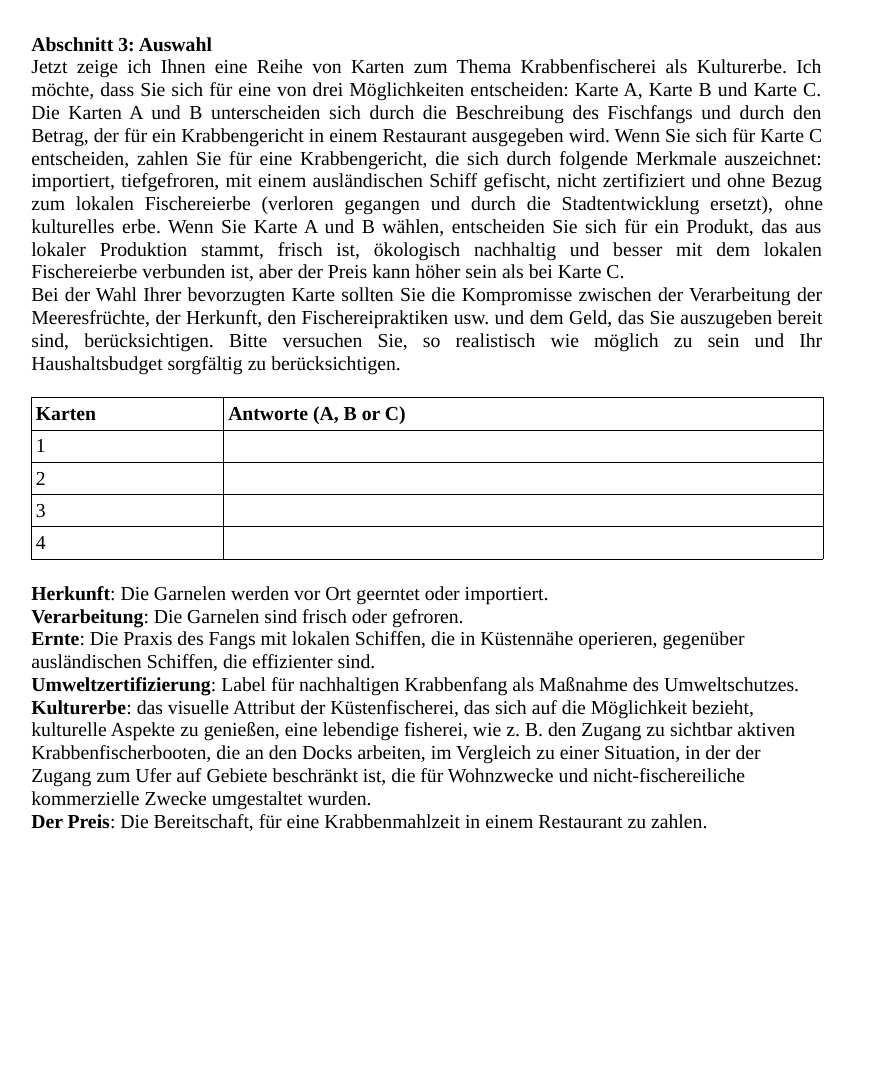}
\end{figure}

\begin{figure}[!htbp]
\centering
  \includegraphics[width=1\textwidth]{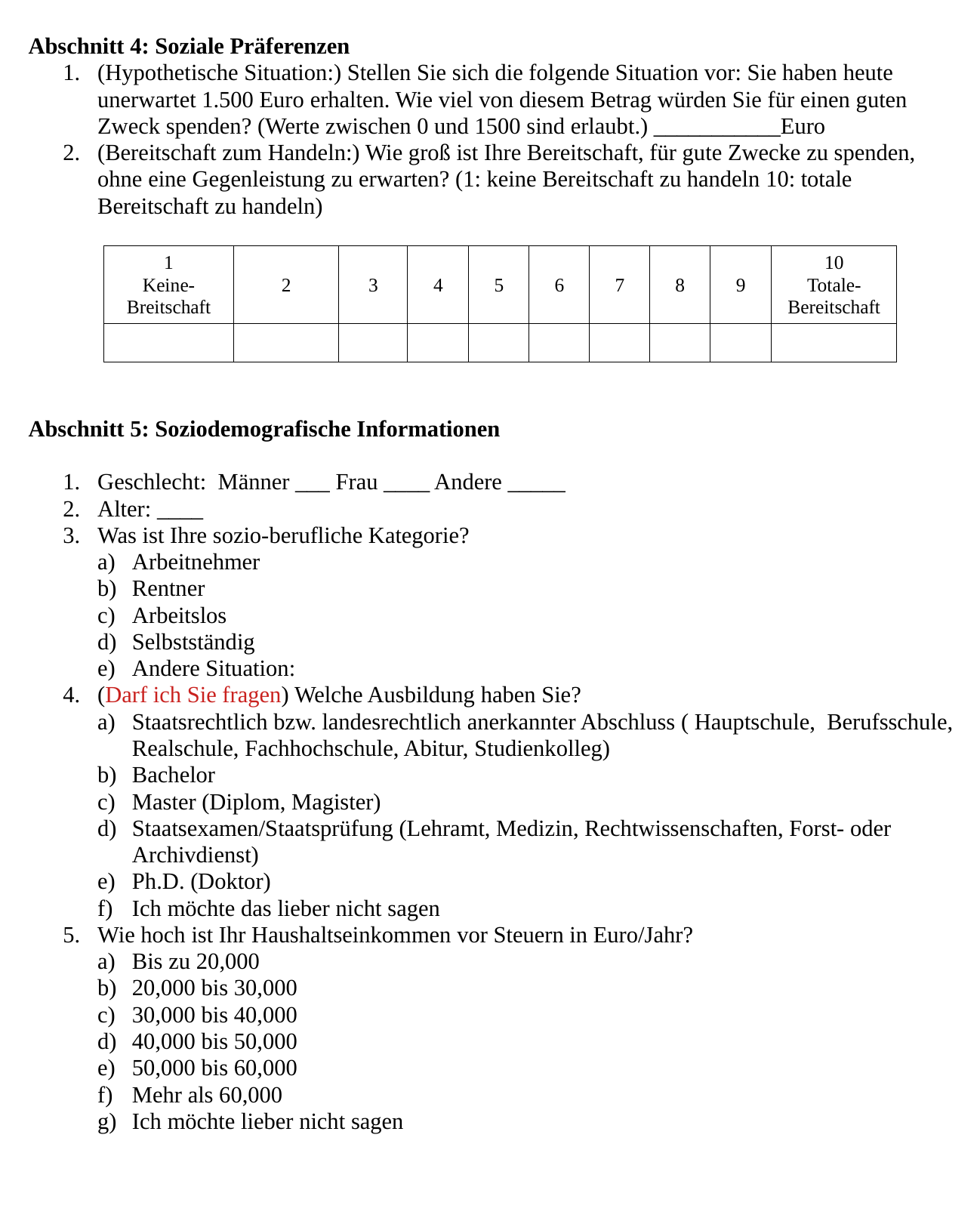}
\end{figure}

%\begin{figure}[!htbp]
%\centering
%  \includegraphics[width=1\textwidth]{img/survey7.eps}
%\end{figure}

%\begin{figure}[!htbp]
%\centering
% \begin{minipage}{\textwidth}
%  \includegraphics[width=\linewidth]{img/barPlotGeneralPerTown.eps}
%  \caption[]{} 
%  \end{minipage}
%\end{figure}
\newpage

\subsection{Heterogeneity Analysis}

\begin{table}[ht]
\centering
\resizebox{11cm}{!}{  
\begin{tabular}{@{\extracolsep{5pt}}lcccc} 
\hline
\hline
\\[-1.8ex]
  & \textbf{Pro-Heritage} & \textbf{No Pro-Heritage }& \textbf{Campaign }& \textbf{No-Campaign}\\
\\[-1.8ex]
\hline
\\[-1.8ex]
\textbf{Origin} & 18.347 & 5.923 & 24.511 & 7.194\\
\\[-1.8ex]
%St. dev & 14.607 & 8.427 & 22.354 & 5.403\\
%\\[-1.8ex]
\textbf{Processing }& 9.123 & 5.402 & 9.462 & 6.278\\
\\[-1.8ex]
%St. dev & 0.198 & 13.237 & 18.517 & 2.140\\
%\\[-1.8ex]
\textbf{Harvesting} & 19.344 & 5.099 & 19.685 & 6.624\\
\\[-1.8ex]
%St. dev & -18.826 & 3.215 & 17.082 & 10.317\\
%\\[-1.8ex]
\textbf{Certification} & 19.418 & 9.940 & 21.441 & 11.599\\
\\[-1.8ex]
%St. dev & -20.207 & 8.168 & 19.860 & 11.090\\
%\\[-1.8ex]
\textbf{Heritage} & 8.634 & 1.946 & 9.643 & 2.547\\
\\[-1.8ex]
%St. dev & 13.521 & 7.348 & -0.738 & 10.922\\
%\\[-1.8ex]
\hline
\\[-1.8ex]
Observations & 227.000 & 169.000 & 166.000 & 242.000\\
\\[-1.8ex]
\hline
\hline
\end{tabular}
}
\caption[Mean estimates of WTP for sub-groups]{Mean estimates of WTP for sub-groups (Unit \euro{}) using the RPL model. The first two columns show the WTP with the sub-samples of pro and no-pro heritage groups. The last two columns show the WTP with respondents located in campaign and no-campaign towns.}

\label{tab:resultsSubGroups-proHeritage}
\end{table}

\end{document}